\documentclass[11pt]{article}
\usepackage{amssymb}
\usepackage{latexsym}
\usepackage{amsmath}
\usepackage{amsfonts}
\usepackage[dvips]{graphicx}
\usepackage{graphics}
\usepackage[T1]{fontenc}
\usepackage{color}

\def\vs#1{\vspace{#1 mm} }

\newtheorem{Theorem}{Theorem}[part]

\newtheorem{Proposition}{Proposition}[part]

\newtheorem{Lemma}{Lemma}[part]

\newtheorem{Remark}{Remark}[part]

\makeatletter \@addtoreset{equation}{part}

\@addtoreset{Definition}{section}

\@addtoreset{Theorem}{section}

\@addtoreset{Proposition}{section}

\@addtoreset{Property}{section}

\@addtoreset{Assumption}{section}

\@addtoreset{Corollary}{section}

\@addtoreset{Lemma}{section}

\@addtoreset{Remark}{section}

\@addtoreset{Example}{section}

\def \Frac{\displaystyle\frac}

\def \proof{{\noindent \bf Proof. }}
\def \ep{\hbox{ }\hfill$\Box$}
\def\reff#1{{\rm(\ref{#1})}}

\def\no{\noindent}

\def\={\;=\;}
\def\.{\;.}
\def\be{\begin{eqnarray}}
\def\ee{\end{eqnarray}}
\def\beq{\begin{equation}}
\def\eeq{\end{equation}}

\def\b*{\begin{eqnarray*}}
\def\e*{\end{eqnarray*}}

\def\Esp#1{\mathbb{E}\left[#1\right]}

\def\Var#1{{\rm Var}\left[#1\right]}

\def\1{{\bf 1}}

\def\eps{\varepsilon}

\def \E{\mathbb{E}}

\def \N{\mathbb{N}}
\def \R{\mathbb{R}}

\def\P{\mathbb{P}}

\def\Fc{{\cal F}}

\def\Lc{{\cal L}}
\def\Mc{{\cal M}}

\def\Tc{{\cal T}}

\def\Vc{{\cal V}}
\def\Wc{{\cal W}}

\def \Var{\mathbb{V}{\rm ar}}

\def\varphil{{\varphi_\lambda}}

\def\eps{\varepsilon}

\def\L{{\cal L}}

\def\GdO{O}
\def\PtO{o}

\title{Double Kernel estimation of sensitivities}
\date{}
\author{R. Elie  \\
\small CREST-ENSAE
                 \\\small  elie@ensae.fr
 }

\title{Double Kernel estimation of sensitivities \vspace{3mm}}
\author{
Romuald ELIE\\
CEREMADE, CNRS, UMR 753 \\
Universit{\'e} Paris-Dauphine,\\
and CREST \\
\sf elie@ensae.fr
\vspace{5mm}
}

\usepackage[]{fancyhdr}

\begin{document}

\maketitle

\vspace{-5mm}

\begin{center}

\today
\end{center}

\vspace{10mm}

\begin{abstract}
This paper adresses the general issue of estimating the sensitivity
of the expectation of a random variable with respect to a parameter
characterizing its evolution. In finance for example, the
sensitivities of the price of a contingent claim are called the {\it
Greeks}. A new way of estimating the {Greeks} has been recently
introduced by Elie, Fermanian and Touzi \cite{ours} through a
randomization of the parameter of interest combined with non
parametric estimation techniques. This paper studies another type of
those estimators whose interest is to be closely related to the
score function, which is well known to be the optimal Greek weight.
This estimator relies on the use of two distinct kernel functions
and the main interest of this paper is to provide its asymptotic
properties. Under a little more stringent condition, its rate of
convergence equals the one of those introduced in \cite{ours} and
outperforms the finite differences estimator. In addition to the
technical interest of the proofs, this result is very encouraging in
the dynamic of creating new type of estimators for sensitivities.

\thispagestyle{empty}

\vspace{10mm}


\vspace{5mm}

\end{abstract}

\newpage

 \setcounter{page}{1}

\section{Introduction}

This paper is closely related to the work of Elie, Fermanian and
Touzi \cite{ours} and we will try to follow their notations. Let
$\lambda$ be some given parameter in $\R^d$, and define the function
 \b*
 V^\phi(\lambda)
 &:=&
 \E\left[ \phi\left(Z(\lambda)\right) \right] \,,
 \e*
where $Z(.)$ is a parameterized random variable with values in
$\R^n$ and $\phi:\R^n \rightarrow \R$ is a measurable function.
 A well understood issue is the numerical computation of the function $V^\phi(\lambda)$
 by means of a Monte Carlo procedure for example.
 A more difficult problem consists in approximating the sensitivity of $V^\phi$ with respect to the parameter $\lambda$.
 For some given parameter $\lambda^0$, we denote by $\beta^0$ the expression of interest
 defined by
 \be\label{defbeta}
 \beta^0 &:=& \nabla_\lambda V^{\phi}(\lambda^0) \;=\; \nabla_\lambda\E[\phi(Z(\lambda))]_{\mid\lambda=\lambda^0}
 \ee
 In financial applications, $V^\phi$ interprets as the no-arbitrage price of a contingent claim,
 defined by the payoff $\phi\left(Z(\lambda)\right)$,
 in the context of a complete market with prices measured in terms of the price of the non-risky asset.
 The sensitivities of $V^\phi$ with respect to the parameter $\lambda$ are often called {\it Greeks},
 and their interest to practitioners is now well established.\\

 To our knowledge, as for the computation of those sensitivities,
 mainly three methods are considered.
 They are compared in detail in the survey paper of Kohatsu-Higa and Montero \cite{Kohatsumontero}
 and we just present briefly here their construction and main properties. \\
 First, the finite differences method consists in approximating the
 derivative of the price by its variation in response to a small perturbation $\epsilon$ of the parameter of interest
 $\lambda$~:
 \be\label{EqFD}
 \beta^0
 &\sim&
 \frac{V^{\phi}(\lambda^0 +\eps)-V^{\phi}(\lambda^0)}{\eps}\,.
 \ee
 Given a number of Monte Carlo simulation for the prices, the choice of $\epsilon$ is related to an equilibrium
 between the bias and the variance of the estimator.
 For discontinuous payoff functions $\phi$, this method appears inefficient due to the poor precision of approximation \reff{EqFD}.
 A theoretical study of those estimators is reported in L'Ecuyer and Perron \cite{Perron},
 Detemple, Garcia and Rindisbacher \cite{Detemple} or Milstein and Tretyakov
 \cite{Milstein}.\\
 Second, one can invert the differentiation and the expectation
 operators to obtain the pathwise estimator given by a Monte Carlo
 estimation based on the representation
 \b*
 \beta^0
 &=&
 \Esp{\phi'(Z(\lambda^0)) \nabla_\lambda Z (\lambda^0)}\,.
 \e*
 This method, introduced by Broadie and Glasserman \cite{BroadieGlasserman},  therefore requires a lot of regularity on the payoff function $\phi$
 as well as the computation of the tangent process $\nabla_\lambda Z$ of the underlying.
 Efficient numerical schemes for the implementation of this method can be found in Giles and Glasserman \cite{gilesglasserman}.\\
 Finally, one can compute $\beta^0$ by reporting the differentiation operator
 on the regular distribution of the underlying $Z(\lambda)$.
 Whenever this random variable admits a density $f(\lambda,.)$ with respect to the
 Lebesgue measure, we obtain the so-called likelihood ratio
 estimator based on
 \be\label{EqScore}
 \beta^0 &=& \Esp{\phi(Z(\lambda^0)) s(\lambda^0,Z(\lambda^0))}\,,
 \qquad\mbox{with}\quad
 s:= \frac{\nabla_\lambda f}{f}\,.
 \ee
 The application of this trick in finance has also been introduced
 by Broadie and Glasserman \cite{BroadieGlasserman}. This type of
 representation has been generalized by Fournié, Lasry, Lebuchoux, Lions and Touzi
 \cite{flllt} who studied the properties of the random variables $\pi$ satisfying
 \b*
 \Esp{\phi(Z(\lambda^0)) \pi}\,, \qquad \mbox{for any function
 } \phi\in\L^\infty(\R^n,\R)\,.
 \e*
 By means of a Malliavin integration-by-parts argument,
 they characterized the set of the so called greek weights $\pi$,
 allowing their tedious computation in some particular cases.
 Nevertheless, beyond all those greek weight based estimators,
 the one related to the score function $s$ and given by \reff{EqScore}
 leads to the smallest variance.\\

 As in \cite{ours}, the main purpose of this paper is to study estimators of the Greek $\beta^0$ whenever the payoff function lacks regularity and the density $f$ of the underlying is unknown. 
 As detailed in the next section, a randomization of the parameter of interest $\lambda$ allows to rewrite the sensitivity $\beta^0$ given by \reff{EqScore} as a conditional  expectation. 
 Combining a non parametric estimation of this conditional expectation with a truncation argument and a kernel estimation of the unknown score function $s$ leads to our estimator $\tilde\beta_n$.
 A slightly different form of $\tilde\beta_n$, without the useful truncation modification, is presented in \cite{ours}, where it serves as a basis to introduce other ones through an integration by part argument. The main contribution of this paper is the presentation of the rather demanding derivation of its asymptotic properties suggested in \cite{ours}. The use of a truncated version of the classical kernel estimator allows to reduce the induced required assumptions on the coefficients. We provide the asymptotic mean square error and distribution of the proposed estimator, leading to the common calibration of the different parameters of simulation.\\

 Despite the more general form of $\tilde\beta_n$, it surprisingly achieves the same rate of convergence rate as the one introduced in \cite{ours}. 
 From a practical perspective, we have to admit that, as argued in \cite{ours}, its numerical implementation is more demanding. Nevertheless, the choice of the two distinct Kernel functions increases significantly the class of possible sensitivity estimators.
  From a technical point of view, the asymptotics of the estimator require a precise derivation of the properties of a kernel estimator of the score function, which appear to be of great interest in themselves. Therefore, this paper offers a new contribution to the literature of the combination of several non-parametric estimators, and
 its particular application to the computation of the Greeks is furthermore promising in the development of competitive numerical computation of sensitivities.\\


 The paper is organized as follows. In section \ref{Section2}, we present in detail the construction of this estimator.
 Its asymptotic properties as well as its practical implementation are discussed in Section \ref{sectasymptotic}.
 Finally, for ease of presentation, the proofs are reported in the last section.




\section{Construction of the estimator}
\label{Section2}

Throughout this paper, we consider a complete probability space
 $(\Omega,\Fc,P)$ supporting a Brownian Motion $W$ valued in
 $\R^m$. We assume that $\Fc$ is the $P$-completion of the
 $\sigma$-algebra generated by $W$. Let $Z(\lambda)$ be a given random variable valued in $\R^n$ and parameterized by $\lambda\in\R^d$
 and $\phi\in\L^\infty(\R^n,\R)$ be a payoff function .
 The purpose of this paper is to construct an estimator of $\beta^0$
 defined in \reff{defbeta} as the sensitivity of $V^\phi$ with respect to $\lambda$ at a given
 point $\lambda^0$.\\

 We shall demonstrate in this section the intuition behind the
 construction of the suggested estimator. We first identify the
 score function $s$ defined in \reff{EqScore} as the optimal Greek weight in the sens of
 \cite{flllt}.
 Considering the realistic case where the score function is unknown, we propose to approximate it through a kernel estimation procedure.
 Combining Monte Carlo simulations with the randomization of the parameter $\lambda$, we are able to
 construct a non-parametric estimator of the score function leading naturally to the estimation of $\beta^0$.
 The reader interested by the asymptotic properties of the estimator should report directly to the next section.

\subsection{The score function as the optimal Greek weight}

We assume that the distribution of $Z(\lambda)$ is absolutely
continuous with respect to the Lebesgue measure, and denote by
$f(\lambda,.)$ the associated density. As announced in the
introduction, under mild smoothness assumptions on the density $f$,
we directly compute that
 \be\label{EqScore2}
 \beta^0 &=& \Esp{\phi[Z(\lambda^0)] s[\lambda^0,Z(\lambda^0)]}\,,
 \qquad\mbox{with}\quad
 s\,:=\, \frac{\nabla_\lambda f}{f} \,=\, \nabla_\lambda \ln{f} \,.
 \ee
 In the context of the Black Scholes model, Broadie and Glasserman \cite{BroadieGlasserman} noticed that this representation
 allows to compute $\beta^0$ by a direct Monte Carlo procedure.
 It is important to notice that the score function $s$ only depends on the distribution of the underlying $Z(\lambda^0)$.
 In a more general framework, Fournie, Lasry, Lebuchoux, Lions and Touzi \cite{flllt} considered the set
 \b*
 \Wc
 &:=&
 \left\{ \pi\in \Lc^2(\Omega,\R^d)~:~\nabla_\lambda V^\phi(\lambda^0)
                              =E\left[\phi(Z^0) \pi\right]
                              ~\mbox{for all}
                              ~\phi\in\L^\infty(\R^n,\R)
                              \right\}\,.
 \e*
 Assuming that $\E \left|s[\lambda^0,Z(\lambda^0)]\right|^2<\infty$, we
 already notice that $s[\lambda^0,Z(\lambda^0)]\in \Wc$. In \cite{flllt},
 the authors construct a new characterization of the set $\Wc$ by means
 of a Malliavin integration by parts argument. After rather tedious computations,
 this representation allows sometimes to produce some
 alternative Greek weights $\pi$ to the score
 $s[\lambda^0,Z(\lambda^0)]$.
 When the density $f$ and therefore the score function $s$ of the underlying are unknown,
 those alternative weights appear to be very helpful.\\

 Nevertheless, their obtention is unfortunately still limited to particular cases and the following argument
 demonstrate that the estimator based on the score $s[\lambda^0,Z(\lambda^0)]$ is of minimal variance
 beyond the class of Greek weight based estimators. Indeed, from the arbitrariness of
 $\phi$ $\in$ $\L^\infty(\R^n,\R)$, we rewrite
 \b*
 \Wc
 &=&
 \left\{ \pi\in\L^2(\Omega,\R^d)~:~E[\pi|Z(\lambda^0)]=s[\lambda^0,Z(\lambda^0)] \right\} \,.
 \e*
 We then deduce that, for any $\pi\in \Wc$,
 \b*
 \Var\left[\phi[Z(\lambda^0)] \pi\right]
 &=&
 E\left[\phi[Z(\lambda^0)]^2 E[\pi\pi'|Z(\lambda^0)]\right]
 -\nabla V^\phi(\lambda^0)\nabla V^\phi(\lambda^0)' \\
 &\ge&
 E\left[\phi[Z(\lambda^0)]^2 E[\pi|Z(\lambda^0)]E[\pi|Z(\lambda^0)]'\right]
 -\nabla V^\phi(\lambda^0)\nabla V^\phi(\lambda^0)' \\
 &=&
 E\left[\phi[Z(\lambda^0)]^2 s[\lambda^0,Z(\lambda^0)]s[\lambda^0,Z(\lambda^0)]'\right]
 -\nabla V^\phi(\lambda^0)\nabla V^\phi(\lambda^0)' \\
 &=&
 \Var\left[\phi[Z(\lambda^0)] s[\lambda^0,Z(\lambda^0)]\right]\,,
 \e*
 where $'$ denotes the transposition operator. Hence
 \b*
 s[\lambda^0,Z(\lambda^0)]\in\Wc
 &\mbox{is a minimizer of}&
 \Var\left[\phi[Z(\lambda^0)] \pi\right]~,~~\pi\in\Wc \,.
 \e*
 As in \cite{ours}, we intend in this paper to construct a non parametric
 estimator based on the approximation of the optimal Greek weight
 given by the unknown score $s[\lambda^0,Z(\lambda^0)]$.

\subsection{Randomization of the parameter}

In order to be able to estimate the unknown score function $s$, the
idea is to create an artificial density around the parameter
$\lambda^0$, on which we can report the differentiation operation.
This well known technique in the non-parametric statistics
litterature, see eg \cite{Sahalia}, is based on the randomization of
the parameter of interest $\lambda$. One may for example interpret
the classical finite difference operator \reff{EqFD} as a particular
case of a randomizing distribution of $\lambda$ with two dirac
masses at
points $\lambda^0$ and $\lambda^0+\eps$.\\

We then introduce $\ell$~: $\R^d\longrightarrow\R$ some given
probability density function, with support containing the origin in
its interior and set
 \b*
 \varphi(\lambda,z)
 \;:=\;
 \ell(\lambda^0-\lambda)\; f(\lambda,z)
 &\mbox{ for}&
 \lambda\in\R^d~\mbox{and}~z\in\R^n \,.
 \e*
 Considering a couple of random variable $(\Lambda,Z)$ with density
 $\varphi$, we therefore rewrite $\beta^0$ as
 \be\label{EqScore3}
 \beta^0 &=& \Esp{\phi(Z) s(\Lambda,Z) \,|\,\Lambda=\lambda^0}\,.
 \ee

 Although we restrict to the case where the density $f$ of the
 underlying $Z(\lambda)$ is unknown, we still consider that we can simulate
 $Z(\lambda)$. This not a limitation in practice since $Z(\lambda)$
 is typically characterized by a stochastic differential equation, which can be classically discretized.
 Hence, we introduce a sequence
 \be \label{observations}
 (\Lambda_i,Z_i)_{1\leq i\le N}
 &\mbox{of $N$ independent r.v. with distribution}&
 \varphi \,,
 \ee
so that, for any $i \leq N$, $\ell(\lambda^0-.)$ is the density of
$\Lambda^i$ and $f(\Lambda^i,.)$ is the conditional density of $Z^i$
given $\Lambda^i$.\\

 We now introduce a kernel function $K:~\R^d \;\rightarrow\; \R$, i.e. such that $\int_{R^d}K=1$.
 Given the $N$ observations $(\Lambda_i,Z_i)_{1\leq i\le N}$, the conditional expectation given by \reff{EqScore3} can be approximated by the classical kernel estimator
 \be\label{barbetaN}
 \bar\beta_N
 &:=&
 \frac{1}{ \ell(0)\, Nh^d }
 \sum_{i=1}^N \phi(Z_i)\;s(\Lambda_i,Z_i)\;
              K\left(\frac{\lambda^0-\Lambda_i}{h}\right)
 \,,
 \ee
 where the bandwidth $h>0$ of the estimator is a small parameter.\\

 This estimator is of course not implementable since the score function $s$ is unknown.
 Nevertheless, as detailed in the next paragraph, the extra regular source of randomness introduced by $\ell$
 allows us to approximate $s$ and leads to a computable estimator of $\beta^0$.

\subsection{The double kernel based estimator} \label{subsect_defnoyaudouble}

 In order to approximate the score function $s$, we shall first estimate the unknown density $\varphi$ of $(\Lambda,Z)$.
 For this purpose, we introduce  a second kernel function $H:~\R^n\;\rightarrow\; \R$.
 Given $N-1$ observations $(\Lambda_j,Z_j)_{1\leq j\le N\,, j\ne i}$,
 we define $\hat\varphi^{-i}$ the classical non-parametric estimator of the density $\varphi$ given by
 \be\label{bandband}
 \hat\varphi^{-i}(\lambda,z)
 &:= &
 \frac{h^{-d-n}}{N-1}\;
 \sum_{j=1,j\neq i}^N
           K\left(\frac{\lambda-\Lambda_j}{h}\right)
           H\left(\frac{z-Z_j}{h}\right)\,.
 \ee
 We denote $\hat\varphil^{-i}(\lambda,z)$ the derivative of this
 estimator with respect to $\lambda$ and we deduce
 \b*
 \hat\varphil^{-i}(\lambda,z)
 &:=&
 \nabla_\lambda\hat\varphi^{-i}(\lambda,z)
 =
 \frac{h^{-d-n-1}}{N-1}\;
 \sum_{j=1,j\neq i}^N
       \nabla K\left(\frac{\lambda-\Lambda_j}{h}\right)
       H\left(\frac{z-Z_j}{h}\right) .\qquad
 \e*

 Observe now that $s$ and $\varphi$ are closely related since we
 easily compute
 \b*
 s(\lambda,z) = \frac{\nabla_\lambda f}{f}(\lambda,z) =  \frac{\nabla_\lambda
 \varphi}{\varphi}(\lambda,z) -
 \frac{\nabla\ell}{\ell}(\lambda^0-\lambda)\,,
  &\mbox{ for}&
 \lambda\in\R^d~\mbox{and}~z\in\R^n \,.
 \e*

 Given the observations $(\Lambda_j,Z_j)_{1\leq j\le N\,, j\ne i}$, this naturally leads to the following estimator $\hat s^{-i}_N$ of the score
 function $s$ given by
 \be\label{snewkernel}
 \hat s^{-i}_N(\lambda,z)
 &:=&
 \frac{\hat\varphil^{-i}}
      {\hat\varphi^{-i}+(\delta/3-\hat\varphi^{-i})\1_{|\hat\varphi^{-i}|<\delta/3}} (\lambda,z)
 \;+\;
 \frac{\nabla\ell(\lambda^0-\lambda)}{\ell(\lambda^0-\lambda)}\, ,
 \ee
 with $\delta$ some small fixed parameter ensuring that the estimator $\hat\varphi^{-i}$ stays away from zero.
 This technical truncation will simply ensure the non explosion of
 the estimator, and the convergence of the estimator will necessitate some
 control on the small values of the true density $\varphi$ detailed
 in Assumption {\rm S} below.\\

 In order to construct an estimator of $\beta^0$, we now replace in \reff{barbetaN}
 each score $s(\Lambda_i,Z_i)$ by the approximation $\hat s^{-i}_N(\Lambda_i,Z_i)$ based on the
 $N-1$ remaining observation. Our estimator is thus defined by
 \be\label{tildebeta}
 \tilde \beta_N
 &:=&
 \frac{1}{ \ell(0)\, Nh^d }
 \sum_{i=1}^N \phi(Z_i)\; \hat s^{-i}_N(\Lambda_i,Z_i)\;
              K\left(\frac{\lambda^0-\Lambda_i}{h}\right)
 \,.
 \ee

 Based on this type of representation, Elie, Fermanian and Touzi \cite{ours} introduce two other estimators by means of an integration by parts argument. 
 Even if the representations proposed in \cite{ours} appears more simple, we surprisingly show in the next section  that our estimator \reff{tildebeta} achieves a similar rate of convergence, under a few more stringent conditions.
 Even if the practical implementation and computation of $\tilde \beta_N$ is more time consuming,
 the general form of \reff{snewkernel} offers a large class of possible estimators, related  to different Kernel functions $K$ and $H$.
 Since the rate of convergence of these estimators is similar, we sincerely believe that this result is very encouraging in the dynamic of creating new type of estimator for sensitivities.
 Moreover the technical proof for the convergence of the estimator appears to be of great interest in itself.



\section{Asymptotic properties} \label{sectasymptotic}

 This section presents the main results of the paper.
 We first provide the asymptotic properties of the estimator $\tilde \beta_N$ defined in \reff{tildebeta}.
 In particular, the obtention of the asymptotic mean square error of the estimator leads to the common optimal choice of the number of
 simulations $N$ and the bandwidth $h$ of the two kernel functions $K$ and $H$.

\subsection{Notations}

 Before stating our results, we recall that the order of a kernel function $K:\R^d\rightarrow\R$ is defined as the smallest non zero integer $p$ such
 that there exist some integers $(j_1,\ldots,j_p)$, with $j_k\in\{1,\ldots,d\}$,
 satisfying
 \b*
 \int l_{\alpha_1}\ldots l_{\alpha_r} K(l)dl
 \;=\;0
 &\mbox{for } \, 0<r < p,\, \alpha_k \in \{ 1,\ldots,d\}, \;\mbox{ \rm and}\\
 \int l_{j_1}\ldots l_{j_p} K(l)dl
 \neq 0 .&  &
 \e*
 Typically, if $K$ is the product of $d$ even univariate kernels, then it is (at least) of order $p=2$.

\vspace{5mm}

In the subsequent subsections, the kernel functions $K$ and $H$ will
be respectively of order $p$ and $q$, and we shall use the notations
 \be\label{defxi}
 \xi^p_K[\psi](\lambda,z)
 &:=&\frac{(-1)^p}{p!}
 \sum_{j_1,\ldots,j_p=1}^d
 \left(\int l_{j_1}\ldots l_{j_p} K(l)dl\right)
      \nabla^p_{\lambda_{j_1}\ldots\lambda_{j_p}}\psi(\lambda,z)\;,\\
 \xi^q_H[\psi](\lambda,z)
 &:=&\frac{(-1)^q}{q!}
 \sum_{j_1,\ldots,j_q=1}^d
 \left(\int v_{j_1}\ldots v_{j_q} H(v)dv\right)
      \nabla^q_{z_{j_1}\ldots z_{j_q}}\psi(\lambda,z)\;, \label{defxi2}
 \ee
for every smooth function $\psi$ defined on $\R^d\times\R^n$. We
shall also denote $A^\otimes\,:=\,AA'$ for every matrix $A$, and $C$
denotes a constant whose value may change from line to line.




\subsection{Asymptotic moments and distribution of the estimator}

 We shall work under the following three assumptions concerning respectively the kernels $K$ and $H$, the payoff
 function $\phi$ and the unknown density function $f$.\\

\no {\bf Assumption KH}\quad {\it $K$ and $H$ are the product of
some univariate compactly supported lipschitz kernels with orders
respectively $p$ and $q$, and $\nabla K$ has bounded variation.
 }

\vs4

\no {\bf Assumption S}\quad {\it $\phi$ is continuous and has
compact support. Moreover, there exist $\delta>0$ such that, for
every $z\in\R^n$, $\inf\left\{\varphi(\lambda,z)~:
            ~(\lambda,z)\in \Vc(\lambda^0)\times C_\phi
\right\} > \delta$, for some neighborhood $\Vc(\lambda^0)$ of
$\lambda^0$, and some compact subset $C_\phi$ of $\R^n$ with ${\rm
Supp}(\phi)\subset{\rm int}(C_\phi)$.
 }

\vs4

\no {\bf Assumption R}\quad {\it For every $\lambda$, the function
$\nabla_\lambda f(\lambda,\cdot)$ is $q$ times differentiable, and
for every integer $j\le q$, the function $\lambda$ $\longmapsto$
$\nabla^j_z\nabla_\lambda \varphi(\lambda,z)$ is continuous at
$\lambda=\lambda^0$ uniformly with respect to $z\in S$, for some
subset $S$ s.t. ${\rm Supp}(\phi)\subset {\rm int}(S)$.  }

 \no {\bf
\textcolor{white}{Assumption R}}\quad {\it For every $z$, the
functions $f(\cdot,z)$ and $\ell$ are $p+1$ times differentiable,
and for every integer $i\le p+1$, the function $\lambda$
$\longmapsto$ $\nabla^i_\lambda f(\lambda,z)$ is continuous at
$\lambda^0$ uniformly with respects to $z\in S$, for some subset $S$
s.t. ${\rm Supp}(\phi)\subset {\rm int}(S)$.  }

\vs4

\begin{Remark}
{\rm We have to admit that {Assumption S} is at first glance rather restrictive on the class of possible payoff functions for financial applications. Nevertheless, we observe that most of the classical ones can be included. In particular, the call option can be considered here even if the payoff does not have compact support. One just need to approximate the greeks associated to the associated Put option and use the correspondence provided by the Call-Put parity relation satisfied in any arbitrage free market.}
\end{Remark}

 We first present the asymptotic bias and variance of the estimator.

\begin{Proposition}\label{proptildebeta}
Under Assumptions {\rm KH, S} and {\rm R}, choose $N$ and $h$ so
that
 \be\label{HypDoubleNh}  h
 \;\longrightarrow\; 0  &\mbox{and}&  \frac{(\ln{N})^4}{N\;h^{d+n+n\vee 2}}
 \;\longrightarrow\; 0  \quad \mbox{as}~~N \;\to\; \infty
 \,.
 \ee
Then, the bias and the variance of $\tilde\beta_N$ satisfy
 \be\label{biasvariancetildebeta}
 \E\left[\tilde\beta_N\right]-\beta^0
 \;\sim\;
 C_1 h^p + C_2 h^q + \frac{C_3}{Nh^{d+n+1}}
 &\mbox{and}&
 \Var\left[\tilde\beta_N\right]
 \;\sim\;
 \frac{\tilde\Sigma}{Nh^{d+2}} \,,
 \ee
where
 \b*
 C_1 &:=&
\frac{1}{\ell(0)} \int \left[\,\xi^p_K\left[\,\ell (\lambda^0-.)
f_\lambda \,+\, \varphil \right] \,-\, \frac{\varphil}{\varphi}
\,\xi^p_K\left[\varphi
\right]\,\right](\lambda^0,z) \phi(z)\,dz\\
 C_2 &:=&  \frac{1}{\ell(0)}
           \int\; \left[\,\xi^q_H\left[\varphil \right] \,-\,
\frac{\varphil}{\varphi}\, \xi^q_H\left[\varphi
\right]\,\right](\lambda^0,z)\;
               \phi(z) \, dz\\
 C_3
 &:=&
 \frac{1}{\ell(0)}
 \int\frac{\phi(z)}{\varphi(\lambda^0,z)}K(l_2-l_1)K(l_1)\nabla K(l_1)
 H^2(v) \, dl_1\, dl_2 \, dv\, dz
 \\
 \tilde\Sigma
 &:=&
 \frac{\E[\phi^2(Z^0)]}{\ell(0)}\;
 \int \left\{\int K(l_2-l_1)\nabla K(l_1)\,dl_1\right\}^\otimes\,dl_2
 \,.
 \e*
\end{Proposition}

 We now turn to the asymptotic distribution of the estimator.

\begin{Theorem} \label{thmtildebeta}
{\rm (i)} Under the conditions of Proposition \ref{proptildebeta},
we have
 \b*
 \sqrt{Nh^{d+2}} \left(\tilde\beta_N - \E[\tilde\beta_N] \right)  &\mathop{\stackrel{law}{\longrightarrow}}
 \limits_{N\rightarrow\infty}&
 \mathcal{N} \left(0,\tilde\Sigma \right)\;.
 \e*
\no {\rm (ii)} If in addition $Nh^{d+2+2(p\wedge q)}\rightarrow 0$,
then the bias vanishes and
 \b*
 \sqrt{Nh^{d+2}} \left(\tilde\beta_N - \beta^0 \right)  &\mathop{\stackrel{law}{\longrightarrow}}
 \limits_{N\rightarrow\infty}&
 \mathcal{N} \left(0,\tilde\Sigma\right)\;.
 \e*
\end{Theorem}

The technical proofs of Proposition \ref{proptildebeta} and Theorem
\ref{thmtildebeta} are reported in Section \ref{preuves}.

\begin{Remark}\label{Rkdouble}
{\rm Note that the condition $n<(p\wedge q)+1$ is necessary in order to satisfy \reff{HypDoubleNh} and the condition of (ii). Thus, for
basket derivatives or bermudean options in finance, it is necessary
to consider high-order kernels, which is not a limitation in
practice.}
\end{Remark}

\subsection{Dependence with respect to the price process dynamics}\label{SectSDE}

 One should typically imagine the random variable $Z$ as the terminal value of a price process $X^\lambda$, whose dynamics are given by a parametrized stochastic differential equation of the form:
 \be\label{SDEZ}
 X_0^\lambda = x(\lambda)\,, \;\;\; dX_u^\lambda = \mu(u,\lambda,X_u^\lambda) du + \sigma(u,\lambda,X_u^\lambda) dW_u,\,. 
 \ee
 where $x:\R^d\rightarrow \R^n$, $\mu:[0,T]\times\R^d\times\R^n\rightarrow \R^n$ and $\sigma:[0,T]\times\R^d\times\R^n\rightarrow \Mc^{n,m}_\R$ are deterministic lipschitz functions. 
 In this case, $Z=X^\lambda_T$ can be simulated easily via any time discretization scheme, even if its density $f$ is unknown.\\
 
 We detail in this paragraph how the regularity of $f$ required in Assumption R can be induced from conditions on the coefficients $x$, $\mu$ and $\sigma$. First, the absolute continuity of $X^\lambda_T$ is ensured by the classical uniform ellipticity condition: suppose the matrix $\sigma \sigma^\top$ is symetric, positive and there exists a constant $c_\sigma>1$  such that 
 \be\label{HypoEllip}
   \frac{1}{c_\sigma} I_d(x) \le \sigma \sigma^\top(t,\lambda,x) \le c_\sigma I_d(x)\,\qquad    \forall(t,\lambda,x)\in [0,T]\times\R^d\times\R^n\,.
 \ee
Second, the density $f$ of $X^\lambda_T$ inherits the regularity of the coefficients $x$, $\mu$ and $\sigma$ through the properties of the corresponding transition densities. Following the arguments of Theorem A.2.2 p.478  in \cite{Azencott}, see also Proposition 5.1 in \cite{Gobet}, Assumption R is satisfied whenever \reff{HypoEllip} holds, $\ell$ is of class $C^1$,  $x$ is of class $C^{q+2}$, and the coefficients $\mu$ and $\sigma$ are of class $C^{1}$ in $(t,\lambda,x)$, $C^{p+2}$ in $\lambda$ as well as $C^{q+2}$ in $x$.\\

It is worth noticing that this analysis gives rise to more tractable assumptions for Proposition \ref{proptildebeta} and Theorem \ref{thmtildebeta} in the realistic framework where $Z$ is the terminal value of a price process with dynamics of the form \reff{SDEZ}.




\subsection{Optimal choice of $N$ and $h$ }
\label{subsectoptimalh}

 We investigate in this section the optimal balance between the number of simulations $N$ and the bandwidth $h$.
 As announced in remark \ref{Rkdouble}, we suppose that $n<(p\wedge q)+1$.
 Under this condition and the assumptions of proposition \ref{proptildebeta}, we obtain a simplification in the asymptotic
 expression of the bias and the mean square error of the estimator
 rewrites
 \b*
 {\rm MSE}(\tilde\beta_N)
 &:=&
 \E\left[ | \tilde\beta_N - \beta^0 |^2 \right]
 \;\sim\;
 \frac{{\rm Tr}(\tilde\Sigma) }{Nh^{d+2}} + |C_1|^2 h^{2p} + |C_2|^2 h^{2q} \; .
 \e*
 Minimizing the MSE in $h$, we get the asymptotically optimal bandwidth selector~:
 \be\label{tildeh}
 \tilde h
 &=&  \left(\frac{(d+2)Tr(\tilde\Sigma)}{2(p\wedge q) |C_1\1_{p\le q} + C_2\1_{q\le p}|^2N}\right)^{1/(d+2(p\wedge q)+2)} \,.
 \ee
 Therefore $\tilde h$ is of order $N^{-1/(d+2(p\wedge q)+2)}$, leading to a MSE of order $N^{-2(p\wedge q)/(d+2(p\wedge q)+2)}$.
 Consequently, despite its more complicated form, the double kernel estimator achieves the same rate of convergence as the one introduced in \cite{ours}.
 The only constraint is the use of kernel functions of order sufficiently large, i.e. satisfying $p\wedge q >n-1$.
 Since, given a large number of simulations, one should always use a kernel function of high order, this constraint is not relevant in
 practice.

\subsection{Remarks and extensions}
\label{subsectextens}

 In this section, we regroup some remarks and possible extensions of the method, which unfortunately go beyond the scope of the paper.\\[-1mm]

 Considering a randomizing distribution $\ell$ with radius equal to the bandwidth $h$, we can improve the rate of convergence of the estimator.
 Indeed, the asymptotic variance of the estimator then reduces to a term of order $1/\sqrt{Nh^2}$, leading to a MSE of order $N^{-(p\wedge q)/(p\wedge q)+1}$. Remarkably, the speed of convergence of the estimator does not depend in this case on the dimension of the underlying $X$.
 For a continuous payoff function, the best finite differences estimator achieves an MSE of order $N^{-4/5}$, see \cite{Detemple}.
 Therefore this estimator outperforms the finite differences one as soon as $p\wedge q > 4 \vee (n-1)$.
 We choose to omit the proof of this result which is technically rather
 demanding.\\[-1mm]
 
 With no doubt, the choice of the randomizing function $\ell$ is crucial for the precision of the estimator presented here. In the particular case of a uniform randomizing distribution $\ell$, the analytical form of the estimator simplifies and, after tedious asymptotic developments, we can see that the optimal choice for the radius of the distribution $\ell$ is the bandwidth $h$ of the kernel function $K$, i.e. the particular case discussed above. From an empirical point of view, the optimal choice of the randomizing density $\ell$ should be intimately related to the choice of the Kernel function $K$. A simple example where these two density functions are identical can naturally be considered.\\

 As for the practical calibration of the optimal bandwidth $\tilde h$ given by \reff{tildeh},
 we need to estimate the constants $C_1$, $C_2$ and $\tilde\Sigma$.
 As for the choice of the bumping parameter of the finite differences estimator, they can be approximated by a preliminary Monte Carlo procedure with very few simulations.
 For example, the procedure proposed in \cite{ours}, can be directly
 adapted to this setting.\\[-1mm]

Finally, a generalization of the above estimator could be considered by taking two different bandwidths.
 Intuitively, the bandwidth for the estimation of the score function introduced in \reff{bandband} should be
 smaller than the one considered for the approximation of the conditional expectation in \reff{barbetaN}.
 Indeed, the signification of those two parameters are rather different, but this question is left for further research.




\section{Proofs}

\label{preuves}

This section is dedicated to the proof of Proposition
\ref{proptildebeta} and Theorem \ref{thmtildebeta}, characterizing
the asymptotic behavior of $\tilde\beta_N$. In this section, we
shall always work under the Assumptions of Proposition
\ref{proptildebeta}.

\subsection{Preliminaries}

Recall that
 \be\label{recalltildebeta}
 \tilde \beta_N
 &:=&
 \frac{1}{ \ell(0)\, Nh^d }
 \sum_{i=1}^N \phi(Z_i)\; \hat s^{-i}_N(\Lambda_i,Z_i)\;
              K\left(\frac{\lambda^0-\Lambda_i}{h}\right)
 \,,
 \ee
where
 \b*
 \hat s^{-i}_N(\lambda,z)
 \;:=\;
 \frac{\hat\varphil^{-i}}
      {\hat\varphi^{-i,\delta}}(\lambda,z)
 \;+\;
 \frac{\nabla\ell}{\ell}(\lambda^0-\lambda)\,,
 \e*
with $\hat\varphi^{-i,\delta}:=\hat\varphi^{-i} +
(\delta/3-\hat\varphi^{-i}) \1_{|\hat\varphi^{-i}|\le\delta/3}$ a
truncated version of $\hat\varphi^{-i}(\lambda,z)$ defined by
 \b*
 \hat\varphi^{-i}(\lambda,z)
 \;:=\;
 \frac{h^{-d-n}}{N-1}\;
 \sum_{j=1,j\neq i}^N
           K\left(\frac{\lambda-\Lambda_j}{h}\right)
           H\left(\frac{z-Z_j}{h}\right)
 \quad{\rm and}\quad\hat\varphil^{-i}
 \;=\;
 \nabla_\lambda\hat\varphi^{-i}\,.
 \e*
For every $\lambda,z$, we set \b*
\bar\varphi(\lambda,z)&\;:=\;&E[\hat\varphi^{-1}(\lambda,z)]
 =
 \int K(l) H(v) \varphi(\lambda-hl,z-hv)\, dl\, dv\,,
 \e*
and its derivative is given by
 \b*
\bar\varphi_\lambda(\lambda,z)
 &=&
h^{-1} \int \nabla K(l) H(v) \varphi(\lambda-hl,z-hv)\, dl\, dv \e*

 Arguing as in the proof of Proposition 4.1 in \cite{ours}, a Taylor expansion combined with a classical change of variable leads to
\be\label{expansionvarphi} \bar\varphi(\lambda,z) -
 \varphi(\lambda,z) \;=\;
 \xi^p_K[\varphi](\lambda,z)\; h^p
 \;+\; \xi^q_H[\varphi](\lambda,z)\; h^q \;+\;\PtO(h^{p\wedge
 q}).
 \ee

Similarly, we get \be\label{expansionvarphil}
\bar\varphi_\lambda(\lambda,z) - \varphil(\lambda,z) \;=\;
\xi^p_K[\varphil](\lambda,z)\; h^p
 \;+\; \xi^q_H[\varphil](\lambda,z)\; h^q \;+\;\PtO(h^{p\wedge
 q})\,.
 \ee

\begin{Remark}
{\rm Since $\phi$ and $K$ have compact support by Assumption {\rm
S}, it follows that, for sufficiently small $h$, the sum in
\reff{recalltildebeta} is restricted to pairs $(\Lambda_i,Z_i)$ with
values in $C_K\times C_\phi$ where $C_K\subset\Vc(\lambda^0)$ is
defined in Assumption S, and $C_\phi$ is a compact subset of
$\,\mathbb{R}^n$ such that ${\rm Supp}\,\phi\subset C_\phi$.}
\end{Remark}

For any function $\psi$ defined on $C_K\times C_\phi$, we set
$$||\psi||_{\infty}\;:=\;\sup_{(\lambda,z)\in C_K\times
C_\phi}|\psi(\lambda,z)|\,,$$
 and, in the following, $||.||_{r}$ refers to the
 $\Lc_r(\Omega)$-norm.

\begin{Remark}\label{remRestexpansionvarphi}
{\rm By Assumption {\rm R}, since $(\lambda,z)$ vary in a compact
subset of $\R^d\times\R^n$, the remainder terms in
\reff{expansionvarphi} and \reff{expansionvarphil} are uniformly
bounded in $(\lambda,z)$. By the same argument, we also see that
$\xi_K^p[\varphi]$, $\xi_H^q[\varphi]$, $\xi_K^p[\varphil]$ and
$\xi_H^q[\varphil]$ are uniformly bounded so that~:}
 \be\label{unifborn}
 \left\|\bar\varphi-\varphi\right\|_\infty
 \;=\;
 \GdO\left( h^{p\wedge q}\right)
 &\mbox{and}&
 \left\|\bar\varphil-\varphil\right\|_\infty
 \;=\;
 \GdO\left( h^{p\wedge q}\right) \,.
 \ee
\end{Remark}

 \vspace{0.2cm}

 We now study further the tails of the estimators $\hat\varphi^{-i}$
and we obtain the following estimates.

 \begin{Lemma}
 There exists $\alpha_1$ and $\alpha_2$ such that
 \be\label{ContEcart1}
 \sup_{i\le N}\,\P[|\hat\varphi^{-i}-\bar\varphi|(\lambda,z)>t]
 &\le& 2 e^{-\frac{t^2}{\alpha_1+\alpha_2 t}N h^{d+n}}
 \,,\quad (\lambda,z)\in C_K \times C_\phi\,.
 \ee
 Furthermore, for any $t>0$, there exists $C_t>0$ and $c_t>0$ satisfying
 \be\label{ContEcart2}
 \P\left[\sup_{i\le N}\|\hat\varphi^{-i}-\bar\varphi\|_\infty >t\right]
 &\le& C_t N^3 e^{- c_t N h^{d+n}}
 \,.
 \ee
 Finally, for any integer $r\ge 1$, we have
 \be\label{estimatepollardmoment1}
 \left\|\sup_{1\le i\le N}
 \left\|\hat\varphi^{-i}-\bar\varphi\right\|_\infty\right\|_{2r}
 &=&
 \GdO\left(\frac{\ln(N)}{\sqrt{Nh^{d+n}}}\right)\,.
 \ee
 \end{Lemma}

 \proof Observe first that there exists $\alpha_1$ and $\alpha_2$
 such that, for any $(\lambda,z)\in C_K \times C_\phi$, the random
 variables $K[(\lambda-\Lambda^i)/h]H[(z-z^i)/h]$ are bounded by
 $3\alpha_2/2$ and, by the usual change of variable, their variance
 are bounded from above by $\alpha_1 h^{d+n}/2$. Therefore \reff{ContEcart1}
 follows directly from the Bernstein inequality.

 \vs2

 We now turn to the proof of the second estimate and first observe that
 \be\label{tempoup}
 \P\left[\sup_{i\le N}\|\hat\varphi^{-i}-\bar\varphi\|_\infty>t\right]
 &\le&
 N\,\P[\|\hat\varphi-\bar\varphi\|_\infty>t],
 \ee
 where, for ease of notation in this proof, we introduce $\hat\varphi:=\hat\varphi^{-1}$.
 Applying the Liebscher strategy, see \cite{Liebscher},
 we recover the compact set $C_K\times C_\phi$ by $C_0\, (R_{N,h})^{-d-n}$
 balls $B_j:=B((\lambda_j,z_j),R_{N,h})$, with $C_0$ a constant chosen
 large enough. On each ball $B_j$, we have
 \be\label{Oups}
 \sup_{B_j} |\hat\varphi-\bar\varphi|
 &\le&
 |\hat\varphi-\bar\varphi|(\lambda_j,z_j)
 +
 \sup_{(\lambda,z)\in B_j}
 |\hat\varphi(\lambda,z)-\hat\varphi(\lambda_j,z_j)|\\
 &+&
 \sup_{(\lambda,z)\in B_j}
 |\bar\varphi(\lambda,z)-\bar\varphi(\lambda_j,z_j)|\nonumber
 \ee
 According to Assumption {\rm KH}, the kernel functions $K$ and $H$ are
 lipschitz and compactly supported. Therefore, there exists $M>0$ such
 that
 \b*
 \sup_{(\lambda,z)\in B_j} |\hat\varphi(\lambda,z)-\hat\varphi(\lambda_j,z_j)|
 &\le&
 C \frac{R_{N,h}}{h} \hat\psi(\lambda_j,z_j),
 \e*
 where $\hat\psi$ is the classical histogram kernel estimator of the density
 $\varphi$ defined by
 \b*
 \hat\psi(\lambda,z)
 &:=&
 \frac{1}{4M^2\,Nh^{d+n}}\,\sum_{i=1}^N \1_{|\Lambda_i-\lambda|\le M h}\1_{|Z_i-z|\le M h}
 \,.
 \e*
 Introducing the notation $\bar\psi:=\E[\hat\psi]$ and choosing $R_{N,h}$ such that $R_{N,h}=\PtO(h)$,
 we then deduce from \reff{Oups} that
 \b*
 \sup_{B_j} |\hat\varphi-\bar\varphi|
 &\le&
 |\hat\varphi-\bar\varphi|(\lambda_j,z_j)
 +
 |\hat\psi-\bar\psi|(\lambda_j,z_j) + 2 C
 \frac{R_{N,h}}{h}\,\bar\psi(\lambda_j,z_j)\,.
 \e*
 Summing up over all the balls $B_j$, we get
 \b*
 \P[\|\hat\varphi-\bar\varphi\|_\infty>t]
 &\le&
 C_0 R_{N,h}^{-(d+n)}
 \left(
 \P[|\hat\varphi-\bar\varphi|(\lambda_j,z_j)>t/3]
 +
 \P[|\hat\psi-\bar\psi|(\lambda_j,z_j)>t/3]
 \right)\\
 &&+
 C_0 R_{N,h}^{-(d+n)}\,\P[2Ch^{-1}R_{N,h}\,|\bar\psi|(\lambda_j,z_j) > t/3]
 \,.
 \e*
 Therefore, applying estimate \reff{ContEcart1} to both kernel
 estimators $\hat\varphi$ and $\hat\psi$, we deduce the existence
 of $\gamma_1$ and $\gamma_2$ satisfying
 \be\label{Jesaispas}
 \P[\|\hat\varphi-\bar\varphi\|_\infty>t]
 \le
 C R_{N,h}^{-(d+n)}
 \left(
 e^{-\frac{t^2}{\gamma_1+\gamma_2 t}N h^{d+n}}
 + \P\left[2C\frac{R_{N,h}}{h}\,|\bar\psi|(\lambda_j,z_j) > t/3\right]
 \right)\,.\quad
 \ee
 But $\bar \psi$ is bounded so that for any given $t$ the last term
 on the right hand side equals $0$ for $h$ small enough.
 Since $Nh^{d+n}\rightarrow\infty$ according to \reff{HypDoubleNh},
 choosing $R_{N,h}=h^2$, we deduce \reff{ContEcart2} from \reff{tempoup}.

 \vs2

 We now turn to the moment inequalities and introduce the notation
 \b*
 Y_N &:=& \frac{\sqrt{Nh^{d+n}}}{\ln(N)}\,\sup_{i\le
 N}\|\hat\varphi^{-i}-\bar\varphi\|_\infty\,,
 \e*
 so that we simply need to prove $\|Y_N\|_{2r}<\infty$ for all integer $r\ge
 1$. Fix $r\in\N^*$ and observe that
 \be\label{sowhat}
 \Esp{Y_N^{2r}} &=& \int_0^\infty 2r s^{2r-1}\P[Y_N>s]ds \;\le\;
 C_a + \int_a^\infty 2r s^{2r-1}\P[Y_N>s]ds\,,
 \ee
 for any a>0.
 We now fix $s$ large enough and take $R_{N,h}=h\ln(N)/\sqrt{N h^{d+n}}$ in
 \reff{Jesaispas} and \reff{tempoup}, so that we get, for $N$ large enough, the
 existence of $\delta_1$ and $\delta_2$ satisfying
 \b*
 \P[Y_N>s] \,\le\,
  C N \left(\frac{\sqrt{N h^{d+n}}}{h\ln(N)}\right)^{d+n}
 e^{-\frac{s\,\ln(N)^2}{\delta_1+\delta_2 s
 \ln(N)/\sqrt{Nh^{d+n}}}}\,.
 \e*
 Since $\ln(N)/\sqrt{Nh^{d+n}}\rightarrow 0$ and $h\rightarrow 0$, we deduce that for $N$ large enough,
 we have
 \b*
 \P[Y_N>s] \,\le\, C N^{d+n} e^{-\frac{s\,\ln(N)^2}{\delta_1+\delta_2 s
 \ln(N)/\sqrt{Nh^{d+n}}}}
 \,\le\, C e^{(d+n)\ln(N) - s(\ln{N})^{3/2}} \,\le\, C e^{-s}\,.
 \e*
 Plugging this estimate into \reff{sowhat} completes the proof.
 \ep\\

Since $\nabla K$ has bounded variation, the exact same reasoning can
apply to the estimators $\hat\varphi^{-i}_\lambda$ and we similarly
derive
 \be
 \left\|\sup_{1\le i\le N}
 \left\|\hat\varphil^{-i}-\bar\varphil\right\|_\infty\right\|_{2r}
 \,=\,
 \GdO\left(\frac{\ln{N}}{h\sqrt{Nh^{d+n}}}\right)
 \,,\quad r\in\N^*\,.\label{estimatepollardmoment2}
 \ee

 The estimates of the previous lemma also allow to control the error due
 to the truncation of $\hat\varphi^{-i}$. Indeed,
 since the function $\varphi$ admits $\delta$ as a lower bound according to Assumption S,
 it follows from \reff{unifborn} that that $\bar\varphi>2\delta/3$
 for $h$ small enough, and \reff{ContEcart1} leads to
 \be\label{unifborn0}
 \P[|\hat\varphi^{-1}(\lambda,z)|<\delta/3] &\leq&
 \P[|\hat\varphi^{-1}-\bar\varphi|(\lambda,z)>\delta/3] \,\leq\,
 2\, e^{- C Nh^{d+n}}\,.\qquad
 \ee
 Introducing $\bar\varphi^{\delta}:=\Esp{\hat\varphi^{-1,\delta}}$,
 we derive
 \be\label{unifborn00}
 \left\|\bar\varphi^{\delta}-\bar\varphi\right\|_\infty
 &\leq&
 \frac{\delta}{3} \,\sup_{C_K\times C_\phi} \P[|\hat\varphi^{-1}|(\lambda,z)<\delta/3]
 \;\leq\;
 \frac{2\delta}{3}\,e^{- C Nh^{d+n}}\,,\qquad
 \ee
 and combining \reff{HypDoubleNh} and \reff{unifborn}, we deduce
 \be\label{unifborn2}
 \left\|\bar\varphi^{\delta}-\varphi\right\|_\infty
 \;=\;
 \GdO\left( h^{p\wedge q}\right)\,.
 \ee
 Similarly, applying \reff{ContEcart2}, we get
 \be
 \left\|\sup_{1\le i\le N}
 \left\|\hat\varphi^{-i,\delta}-\hat\varphi^{-i}\right\|_\infty\right\|_{2r}
 &\le&
 \delta \, \P\left[\sup_{i\le N}\|\hat\varphi^{-i}-\bar\varphi\|_\infty
 >\delta/3\right]\nonumber\\
 &\le& C \delta N^3 e^{- C Nh^{d+n}}\,,
 \quad r\in\N\,.\label{unifborn7}
 \ee
 Observe also that \reff{unifborn00} and \reff{unifborn7} combined
 with \reff{HypDoubleNh} allows to derive
 \be\label{estimatepollardmoment3}
 \left\|\sup_{1\le i\le N}
 \left\|\hat\varphi^{-i,\delta}-\bar\varphi^\delta\right\|_\infty\right\|_{2r}
 \,=\,
 \GdO\left(\frac{\ln{N}}{\sqrt{Nh^{d+n}}}\right)\,,
 &&\mbox{for any }\; r\in\N^*\,.
 \ee

 \vspace{0.2cm}

Finally, since $(\lambda,z)$ vary in a compact subset, Assumptions
{\rm R} and {\rm S} imply that
 \be\label{upperboundvarphi1}
 \left\|\varphi\right\|_\infty
 \,+\,
 \left\|\varphil\right\|_\infty
 \,+\,
 \left\|1/\varphi\right\|_\infty
 \;<\; \infty\,.
 \ee
It then follows from equation \reff{unifborn}, \reff{unifborn2} and
the truncation procedure that
 \be\label{upperboundvarphi2}
 \left\|\bar\varphi\right\|_\infty
 +
 \left\|\bar\varphi^\delta\right\|_\infty
 +
 \left\|\bar\varphil\right\|_\infty
 +
 \left\|1/\bar\varphi\right\|_\infty
 +
 \left\|1/\bar\varphi^\delta\right\|_\infty
 +
  \sup_{1\le i\le N}\;
 \left\|1/\hat\varphi^{-i,\delta}\right\|_\infty
 &<& \infty\,.\qquad
 \ee

 \vspace{0.2cm}

\subsection{A suitable decomposition}

For any $N\in\N$ and $i\le N$, we define the following functions
$t^1_{i,N},\ldots,t^9_{i,N}$ on
$\mathbb{R}^d\times\mathbb{R}^n\times\Omega$~:
 \b*
 t^1_{i,N}\;:=\; s \,,\;\;
 t^2_{i,N} \;:=\; \frac{\bar\varphil-\varphil}{\varphi} \,,\;\;
 t^3_{i,N} \;:=\; \frac{(\varphi-\bar\varphi^\delta)\,\varphil}{\varphi^2} \,,\;\;
 t^4_{i,N} \;:=\; \frac{(\varphi-\bar\varphi^\delta)\,(\bar\varphil\,\varphi-\bar\varphi^\delta\,\varphil)}{\varphi^2\,\bar\varphi^\delta} \,,\\
 t^5_{i,N} \;:=\; \frac{\hat\varphil^{-i}-\bar\varphil}{\varphi} \,,\;\;
 t^6_{i,N} \;:=\; \frac{(\bar\varphi^\delta-\hat\varphi^{-i,\delta})\,\bar\varphil}{(\bar\varphi^\delta)^2} \,,\;\;
 t^7_{i,N} \;:=\; \frac{(\hat\varphil^{-i}-\bar\varphil)\,(\varphi^\delta-\bar\varphi^\delta)}{\varphi^\delta\,\bar\varphi^\delta}\,,\\
 t^8_{i,N} \;:=\; \frac{(\bar\varphi^\delta-\hat\varphi^{-i,\delta})(\hat\varphil^{-i}-\bar\varphil)}{\hat\varphi^{-i,\delta}\,\bar\varphi^\delta}\,\;\;
 {\rm and} \quad t^9_{i,N} \;:=\;
 \frac{(\bar\varphi^\delta-\hat\varphi^{-i,\delta})^2\bar\varphil}{\hat\varphi^{-i,\delta}\,(\bar\varphi^\delta)^2},
 \e*
 \b*
 \quad {\rm so\,\,that} \qquad
\hat s_N^{-i}(\Lambda_i,Z_i)\,=\,\sum_{j=1}^9
t^j_{i,N}(\Lambda_i,Z_i)\, .
 \e*

This implies the following decomposition of the estimator
$\tilde\beta_N$~:
 \be\label{decompotildebeta}
 \tilde\beta_N
 \;=\;
 \sum_{j=1}^9 T^j_N,
 \mbox{ where }
 T^j_N
 \;:=\;
 \frac{1}{\ell(0)\, Nh^{d} }
 \sum_{i=1}^N \phi(Z_i)\; t^j_{i,N}(\Lambda_i,Z_i)\;
                  K\left(\frac{\lambda^0-\Lambda_i}{h}\right)\,,\quad
 \ee
for every $j=1,\ldots,9$. By \reff{upperboundvarphi1} and
\reff{upperboundvarphi2}, we observe that
 \b*
 \left\|t_{i,N}^j\right\|_\infty \,<\, \infty\,,
 &\mbox{ for all}&
 j=1,\ldots,4 \,.
 \e*

\begin{Lemma}\label{lemExpectation}
For any $j=1,\ldots,4$, we have $\E\left[T^j_N\right]$ $=$
${\GdO}\left(\left\|t_{1,N}^j\right\|_\infty\right)$.
\end{Lemma}

\proof The result is derived from the following inequality:
\b*\left|\E[T_N^j]\right| &\leq&
 \frac{1}{\ell(0)\, h^{d} }
\left|\E\left[\phi(Z_1)\; t^j_{1,N}(\Lambda_1,Z_1)\;
                  K\left(\frac{\lambda^0-\Lambda_1}{h}\right)\right]\right|\\
 &\leq&
 \frac{1}{\ell(0)}
 \left|\int\ \phi(z)\; t^j_{1,N}(\lambda^0-hl,z)\;
                  K(l)\,dl\,dv\right|\\
 &\leq&
 C\,||t^j_{1,N}||_{\infty} \,.
\e*\ep

\begin{Lemma}\label{lemVariance}
For every $j=1,\ldots, 4$,
$\Var(T^j_N)={\GdO}\left(N^{-1}h^{-d}\right\|t^j_{1,N}\left\|_\infty^2\right)\,$.
\end{Lemma}

\proof For any $j=,1\ldots, 4$, the $N$ random variables
$T_N^j(\Lambda_i,Z_i)$ are independent and
 \b*
 \Var[T_N^j] &=&
 \frac{1}{\ell(0)^2\,N h^{2d}}
 \Var\left[\phi(Z_1)\; t^j_{1,N}(\Lambda_1,Z_1)\;
                  K\left(\frac{\lambda^0-\Lambda_1}{h}\right)\right]\\
 &\leq&
 \frac{1}{\ell(0)^2\,N h^{2d}}
 \E\left[\phi^2(Z_1)\; t^j_{1,N}(\Lambda_1,Z_1)^2\;
                  K^2\left(\frac{\lambda^0-\Lambda_1}{h}\right)\right]\\
 &\leq&
 \frac{\|t^j_{1,N}\|^2_{\infty}}{\ell(0)^2\,N h^{d}}
 \int\phi^2(z)\; K^2(l)\,dl\,dv
 \,.\,
 \e*\ep

The analysis of $T^j_N$, for $j>4$, requires more effort because of
the dependence between the random variables
$t^j_{i,N}(\Lambda_i,Z_i)$.

\begin{Lemma}\label{lemT5}
$\E[T_N^5]\,=\,0\,$ and $\Var(T^5_N)\;\sim\;\tilde\Sigma/(Nh^{d+2})$
where $\tilde\Sigma$ is defined in Proposition \ref{proptildebeta}.
\end{Lemma}

\proof We introduce for any $i=1,\ldots, N$ and $j=1,\ldots, N$~:
\b*\Tc_{ij}:=\frac{ \phi(Z_i)}{ \varphi(\Lambda_i,Z_i)}\;
              K\left(\frac{\lambda^0-\Lambda_i}{h}\right) \left\{
              \nabla_{\lambda} K \left(\frac{\Lambda_i-\Lambda_j}{h} \right)
              H \left(\frac{Z_i-Z_j}{h} \right)
              - h^{d+n+1}
              \bar\varphil(\Lambda_i,Z_i) \right\}\,,\e*
so that $T^5_N$ can be re-written in
 \b*
 T^5_N\,=\,\frac{h^{-2d-n-1}}{ \ell(0)\,N(N-1)}\sum_{i<j}
 \left( \Tc_{ij}+\Tc_{ji}\right) \;.
 \e*
By definition, for any $i=1,\ldots, N$ and $j=1,\ldots, N$ with
$i\neq j$, we have~
\b*\bar\varphil(\Lambda_i,Z_i)\,=\,\frac{1}{h^{d+n+1}}\E\left[\nabla_{\lambda}
K \left(\frac{\Lambda_i-\Lambda_j}{h} \right)
              H \left(\frac{Z_i-Z_j}{h}\right)\,\mid\,\Lambda_i,Z_i\right].\e*
Therefore, $\E[\Tc_{ij}]=0$ whenever $i\neq j$, leading to
$\E[T^5_N]=0.$

Since the $\Tc_{ij}$ are not independent, the computation of the
variance requires to decompose $T^5_N$ into
 \be\label{T51+T52}
 T^5_N &=& T^{5,1}_N+ T^{5,2}_N \,,
 \ee
where
 \b*
 T^{5,1}_N &:=& \frac{h^{-2d-n-1}}{
 \ell(0)\,N(N-1)}\sum_{i<j} \left(
 \Tc_{ij}+\Tc_{ji}-b(\Lambda_i,Z_i)-b(\Lambda_j,Z_j)\right)\,,\\
 T^{5,2}_N &:=& \frac{h^{-2d-n-1}}{
 \ell(0)\,N(N-1)}\sum_{i<j} \left(
 b(\Lambda_i,Z_i)+b(\Lambda_j,Z_j)\right) \,.
 \e*
and
 $
 b(\lambda,z)
 \,:=\,
 E\left[\Tc_{12}|\Lambda_2=\lambda,Z_2=z\right]
 $.\\

1. Let first study the term $T^{5,1}_N$.\\ Setting
$\Upsilon_{ij}\,:=\,\Tc_{ij}+\Tc_{ji}-b(\Lambda_i,Z_i)-b(\Lambda_j,Z_j)$,
we derive the key property~:
 \be\label{keyproperty}
 \E[\Upsilon_{ij}|\Lambda_i,Z_i]=\E[
 \Upsilon_{ij}|\Lambda_j,Z_j]=0 \,.
 \ee
Therefore $T_N^{5,1}$ has zero mean and we derive~:
$$ \Var[T_N^{5,1}]\,=\,\frac{h^{-4d-2n-2}}{
\ell(0)^2\,N^2(N-1)^2}\sum_{i<j}\E[\Upsilon_{ij}\Upsilon_{ij}']\,=\,\frac{h^{-4d-2n-2}}{
2\ell(0)^2\,N(N-1)}\E[\Upsilon_{12}\Upsilon_{12}'].$$

By \reff{keyproperty}, we compute~:
 \b*\E[\Upsilon_{12}\,\Upsilon_{12}']
 &=&\,2\,\E[\Tc_{12}\Tc_{12}']+2\,\E[\Tc_{12}\Tc_{21}']-2\E[b^2(\Lambda_1,Z_1)]
 \,.
 \e*
We next estimate that $\left|\E[\Tc_{12}\Tc_{12}']\right|$ is
dominated by
 \b*
 &&\E\left[ \frac{ \phi^2(Z_1)}{
 \varphi^2(\Lambda_1,Z_1)}\;
              K^2\left(\frac{\lambda^0-\Lambda_1}{h}\right)
              \left|\nabla_\lambda K \right|^2 \left(\frac{\Lambda_1-\Lambda_2}{h} \right)
              H^2 \left(\frac{Z_1-Z_2}{h} \right)
              \right]
 \\
 &+&\,h^{2d+n}\int \phi^2(z)\;K^2(l_1)
              |\nabla_\lambda K|^2(l_2)
              H^2(v)
              \Frac{\varphi(\lambda^0-hl_1-hl_2,z-hv)}
                   {\varphi(\lambda^0-hl_1,z)}\,
              dl_1\,dl_2\,dz\,dv \,,
 \e*

by the usual change of variables. Clearly, the first term on the
right hand-side is of order $\GdO(h^{2d+n})$, while the second one
is a $\GdO(h^{3d+2n+2})$ by \reff{upperboundvarphi2}. Similarly, we
have $\E[\Tc_{12}\Tc_{21}']\,=\,\GdO(h^{2d+n})$. Moreover,
$\E[b^2(\Lambda_1,Z_1)]=\GdO(N^{-2}h^{-d-2})$. We deduce that
 \be\label{VarianceT51}
 \Var(T^{5,1}_N)\,=\,\GdO \left(
 \frac{1}{ N^2 h^{2d+n+2}}\right)=o\left(
 \frac{1}{Nh^{2+d}}\right) \,,
 \ee
using the relations between $N$ and $h$ given by \reff{HypDoubleNh}.\\

  2. We next rewrite $T^{5,2}_N$ as
 \b*
 T^{5,2}_N
 &=&
 \frac{h^{-2d-n-1}}{ \ell(0)\,N}
 \sum_{i} b(\Lambda_i,Z_i) \,.
 \e*
By the usual change of variables,
 \b*
 b(\lambda,z)
 &=&
 h^{d+n} \int  \phi(z+hv)\;
 K\left(\frac{\lambda^0-\lambda}{h}-l\right)
 \nabla K (l)H (v) \,dl\,dv \\
 & &
 -\,h^{n+1} \int  \phi(z)\;\bar\varphil(\lambda^0 - hl,z) K (l) \,dl.
 \e*
By direct calculation, it is easily checked that the second term is
negligible. Then, by the usual change of variables, it follows that
 \b*
 & & \E[b(\Lambda_i,Z_i) b(\Lambda_i,Z_i)'] \\
 &\sim&
 h^{3d+2n} \int \left\{\int\phi(z+hv)K(l_2-l_1)\nabla K(l_1) H(v) \,
 dl_1\, dv\right\}^\otimes \varphi(\lambda^0-hl_2,z) \, dl_2\, dz\,.
 \e*
By Assumptions {\rm S} and {\rm R}, we deduce from the dominated
convergence theorem together with the fact that
$\E[b(\Lambda_i,Z_i)]=0$ that
 \be\label{VarianceT52}
 \Var[T^{5,2}_N]
 \,\sim\,
 \frac{1}{Nh^{d+2}}
 \int \phi^2(z) \left\{\int K(l_2-l_1) \nabla K(l_1)
 \, dl_1\,\right\}^\otimes \varphi(\lambda^0,z) \, dl_2\, dz\,.
 \ee
The proof is completed by collecting the estimates
\reff{VarianceT51} and \reff{VarianceT52} into \reff{T51+T52}.
 \ep

\begin{Lemma}\label{lemT6}
$\E[T_N^6]\,=\,\PtO(h^{p\wedge q})\,$ and
$\Var(T^6_N)\;=\;\PtO(N^{-1}h^{-d-2})$.
\end{Lemma}

\proof We decompose $t_{i,N}^6$ into the sum of
 \b*
 t^{6,1}_{i,N} \,:=\,
 \frac{(\bar\varphi-\hat\varphi^{-i})\,\bar\varphil}{(\bar\varphi^\delta)^2}\,,\;\;\;\;
 t^{6,2}_{i,N} \,:=\,\frac{(\hat\varphi^{-i}-\hat\varphi^{-i,\delta})\,\bar\varphil}{(\bar\varphi^\delta)^2}
 \;\;\mbox{ and }\;\;
 t^{6,3}_{i,N} \,:=\, \frac{(\bar\varphi^\delta - \bar\varphi)\,\bar\varphil}{(\bar\varphi^\delta)^2}\,,\;
 \e*
and we study the corresponding $T^{6,1}_N$, $T^{6,2}_N$ and
$T^{6,3}_N$ separately.

\vspace{0.1cm}

\no 1. It can be checked easily that $T^{6,1}_N$ can be dealt with
as $T^5_N$. By the same calculation, we get $\E[T^{6,1}_N]=0$ and
$$ \Var(T_N^{6,1})\sim \frac{h^{-4d-2n}}{ \ell(0)^2\,N^2}\sum_{i}
\Var(\tilde b(\Lambda_i,Z_i)) $$ where $\tilde b(\lambda,z)$ is
given by~:
 \b*
 E\left[ \frac{\phi(Z_i)\varphil(\Lambda_i,Z_i)}{ \varphi(\Lambda_i,Z_i)^2}\;
              K\left(\frac{\lambda^0-\Lambda_i}{h}\right) \left\{
              K \left(\frac{\Lambda_i-\lambda}{h} \right)
              H \left(\frac{Z_i-z}{h} \right)
              - h^{d+n}
              \bar\varphi(\Lambda_i,Z_i) \right\}\right]
 \e*
The variables $\tilde b(\Lambda_i,Z_i)$ have also zero mean and, as
in the proof of Lemma \ref{lemT5}, the usual change of variables
implies that
 \b*
 h^{-3d-2n}\;\Var(\tilde b(\Lambda_i,Z_i))
 &\sim&
 \int \left[G_6(l_2,z)\right]^\otimes \varphi(\lambda^0-hl_2,z) \, dl_2\,
 dz\,,
 \e*\vspace{-0.6cm}
 \b*
 \mbox{with}\;\,
 G_6(l_2,z):=
 \int\phi(z+hv)\frac{\varphil}{\varphi}(\lambda^0+hl_1-hl_2,z+hv)K(l_2-l_1)K(l_1)
 H(v) \, dl_1\, dv.
 \e*
By the continuity and the uniform boundedness of $\phi$ and
$\varphil /\varphi$ implied by Assumptions {\rm S} and {\rm R}, we
derive
 \b*
 \Var(T^{6,1}_n)
 &=&
 \GdO \left(\frac{1}{Nh^d}\right)
 \,=\,\PtO
 \left(\frac{1}{Nh^{d+2}}\right)\,.
  \e*

\vspace{0.1cm}

\no 2. We now turn to $T^{6,2}_N$ and compute
 \b*
 |T^{6,2}_N| \;\leq\;C\,
 \sup_{i\le N}\left\|\hat\varphi^{-i,\delta}-\hat\varphi^{-i}\right\|_\infty
 \left(\frac{1}{Nh^d}\sum_{i=1}^N
 \left|\phi(Z_i)K\left(\frac{\lambda^0-\Lambda_i}{h}\right)\right|\right)\,.
 \e*
 Therefore, we deduce from Cauchy-Schwarz inequality that
 \b*
 \left|\Esp{T^{6,2}_N}\right| \;\leq\;C\,
 \left\|\sup_{i\le
 N}\left\|\hat\varphi^{-i,\delta}-\hat\varphi^{-i}\right\|_\infty\right\|_2
 \Esp{
 \left(\frac{1}{Nh^d}\sum_{i=1}^N
 \left|\phi(Z_i)K\left(\frac{\lambda^0-\Lambda_i}{h}\right)\right|\right)^2}^{1/2}\,,
 \e*
 and \reff{HypDoubleNh} combined with \reff{unifborn7} lead to $ \Esp{T^{6,2}_N} \;=\; \PtO\left(h^{p\wedge
 q}\right)$.
 Similarly, we get
 \b*
 Var(T^{6,2}_N) \;\leq\;C\,
 \left\|\sup_{i\le
 N}\left\|\hat\varphi^{-i,\delta}-\hat\varphi^{-i}\right\|_\infty\right\|_4
 \Esp{
 \left(\frac{1}{Nh^d}\sum_{i=1}^N
 \left|\phi(Z_i)K\left(\frac{\lambda^0-\Lambda_i}{h}\right)\right|\right)^4}^{1/4}\,,
 \e*
 which leads to $\Var(T^{6,2}_n) \,=\,\PtO\left(N^{-1}h^{-d-2}\right)$.

\vspace{0.1cm}

\no 3. We finally observe that $T^{6,3}_N$ is treated similarly
thanks to \reff{unifborn00}. \ep

\begin{Lemma}\label{lemT7}
$\E[T_N^7]\,=\,0\,$ and $\Var(T^7_N)\;=\;\PtO(N^{-1}h^{-d-2})$.
\end{Lemma}

\proof Observe that
 \b*
 t^7_N(\lambda,z)
 \;=\;
 t^5_N(\lambda,z)\psi(\lambda,z)
 &\mbox{where}&
 \psi \;:=\; \frac{\varphi - \bar\varphi^\delta}{\bar\varphi^\delta}\,\cdot
 \e*
Following the lines of the proof of Lemma \ref{lemT5}, we see that
$\E[T^7_N]=0$, and we estimate
 \b*
 Nh^{d+2}\Var(T^7_N)
 &\sim&
 \int \left[G_7(u,z)\right]^\otimes \varphi(\lambda^0-hu,z) \, du\,
 dz\,,
 \e*\vspace{-0.6cm}
 \b*
 \mbox{with }\quad
 G_7(u,z)
 \;:=\;
 \int\phi(z+hv)\psi(\lambda^0+hl-hu,z+hv)K(u-l)\nabla K(l)
 H(v) \, dl\, dv\,.
 \e*
By \reff{unifborn2} and \reff{upperboundvarphi2} it follows that
$\|\psi\|_{\infty}=\GdO(h^{p\wedge q})$ and, since $\varphi$ and
$\phi$ are uniformly bounded, we deduce that
 \b*
 \Var(T^7_N)
 &=&
 \GdO\left(\frac{h^{p\wedge q}}{Nh^{d+2}}\right)
 \;=\;\PtO\left(\frac{1}{Nh^{d+2}}\right) \,.
 \e*
\ep
\begin{Lemma}\label{lemT8}
$\E\left[T^8_N\right] \sim
 \Frac{h^{-d-n-1}}{\ell(0)N}
 \left(\int\phi\right)
 \left(\int H^2\right)
 \int K(l_1-l_2)K(l_2)\nabla K(l_2)dl_1dl_2$\\
and $\Var(T^8_N)\;=\;\PtO(N^{-1}h^{-d-2})$.
\end{Lemma}

\proof We split the proof it two steps.\\
1. We first estimate $\E\left[T^8_N\right]$. We rewrite
$t^8_N(\lambda,z)$ as
$t^{8,1}_N(\lambda,z)+t^{8,2}_N(\lambda,z)+t^{8,3}_N(\lambda,z)$
with
 \b*
 t^{8,1}_{i,N}
 &=&
 \frac{(\bar\varphi-\hat\varphi^{-i})(\hat\varphil^{-i}-\bar\varphil)}{\varphi^2}\,,\\
 t^{8,2}_{i,N}
 &=&
  \frac{(\bar\varphi^{\delta}-\bar\varphi)(\hat\varphil^{-i}-\bar\varphil)}{\varphi^2}
  +
 \frac{(\hat\varphi^{-i}-\hat\varphi^{-i,\delta})(\hat\varphil^{-i}-\bar\varphil)}{\varphi^2}\,,\\
 t^{8,3}_{i,N}
 &=&
 \frac{(\bar\varphi^\delta-\hat\varphi^{-i,\delta})^2(\hat\varphil^{-i}-\bar\varphil)}{\hat\varphi^{-i,\delta}\,(\bar\varphi^\delta)^2}
 \,+\,
 \frac{(\bar\varphi^\delta-\hat\varphi^{-i,\delta})(\hat\varphil^{-i}-\bar\varphil)(\varphi^2-(\bar\varphi^\delta)^2)}{\varphi^2\,(\bar\varphi^\delta)^2}\,.
 \e*
Then $T^8_N = T^{8,1}_N+T^{8,2}_N+T^{8,3}_N$, where
 \b*
 T^{8,k}_N
 \;:=\;
 \frac{1}{\ell(0)\, Nh^{d} }
 \sum_{i=1}^N \phi(Z_i)\; t^{8,k}_{i,N}(\Lambda_i,Z_i)\;
                  K\left(\frac{\lambda^0-\Lambda_i}{h}\right)\,,
 &\mbox{ for}& k=1,2,3 \,.
 \e*
We now introduce
 \b*
 \begin{array}{rcl}
 U_{ij}
 &:=&
 \nabla_{\lambda} K\left(\frac{\Lambda_i-\Lambda_j}{h} \right)H \left(\frac{Z_i-Z_j}{h} \right)
 -
 \E\left[\nabla_{\lambda} K \left(\frac{\Lambda_i-\Lambda_j}{h} \right)H \left(\frac{Z_i-Z_j}{h}
 \right)|\Lambda_i,Z_i\right]\,,\\
 V_{ij}
 &:=&
 K\left(\frac{\Lambda_i-\Lambda_j}{h} \right)H \left(\frac{Z_i-Z_j}{h} \right)
 -
 \E\left[K \left(\frac{\Lambda_i-\Lambda_j}{h} \right)H \left(\frac{Z_i-Z_j}{h}
 \right)|\Lambda_i,Z_i\right]\,,
 \end{array}
 \e*
 so that
 \b*
 \E\left[U_{ij}V_{ik}|\Lambda_i,Z_i\right]
 \,=\,
 \E\left[U_{ij}|\Lambda_i,Z_i\right]\E\left[V_{ik}|\Lambda_i,Z_i\right]
 \,=\,
 0 & \mbox{whenever} & j\neq k\,.
 \e*
 Using this property, we compute directly that
 \b*\E\left[t^{8,1}_N(\Lambda_1,Z_1)|\Lambda_1,Z_1\right]
 &=&
 \frac{h^{-2d-2n-1}}{(N-1)^2\varphi^2(\Lambda_1,Z_1)}
 \E\left[\sum_{j\neq 1}\sum_{k\neq 1}U_{1j}\,V_{1k}|\Lambda_1,Z_1\right]\\
 &=&
 \frac{h^{-2d-2n-1}}{(N-1)\varphi^2(\Lambda_1,Z_1)}
 \E\left[U_{12}\,V_{12}|\Lambda_1,Z_1\right]\,.
 \e*
Since the expectation of $T^{8,1}_N$ is given by~:
 \b*
 \E\left[T^{8,1}_N\right]
 &=&
 \frac{h^{-d}}{\ell(0)}
 \E\left[\phi(Z_1)K\left(\frac{\lambda^0-\Lambda_1}{h}\right)
         \E\left[t^{8,1}_{1,N}(\Lambda_1,Z_1)|\Lambda_1,Z_1\right]\right]\,,
 \e*
 we derive by the usual change of variables,
 \b*
 \ell(0)Nh^{d+n+1}\; \E\left[T^{8,1}_N\right]\;
 &\sim&
 \int G_8(l_2,z) \varphi(\lambda^0-hl_2,z) \,dl_2\,dz\,,
 \e*\vspace{-0.6cm}
 \b*
 \mbox{with }\;\;
 G_8(l_2,z)
 :=
 \int\frac{\phi(z+hv)}{\varphi(\lambda^0+hl_1-hl_2,z+hv)}K(l_2-l_1)K(l_1)\nabla K(l_1)
 H^2(v) \, dl_1\, dv \,.
 \e*
Finally, by the continuity and the uniform boundedness of $\varphi$
and $\phi$, we derive~:
 \be\label{estimt81n}
 \E\left[T^{8,1}_N\right]
 &\sim&
 \frac{h^{-d-n-1}}{\ell(0)N}
 \int \phi(z) K(l_2-l_1)K(l_1)\nabla K(l_1)
 H^2(v) \, dl_1\, dv \, dl_2\, dz\,.\qquad
 \ee

Furthermore, by Cauchy-Schwarz inequality and \reff{HypDoubleNh}, we
have
 \be\label{estimestim}
 \left|\E\left[T^{8,k}_N\right]\right|
 &\le&
 \left\| \sup_{i\le N}\left\|t^{8,k}_{i,N}\right\|_\infty\right\|_2
 \E\left[\left(\frac{1}{Nh^d}\sum_{i=1}^N\left|\phi(Z_i)K\left(\frac{\lambda^0-\Lambda_i}{h}\right)\right|\right)^2\right]^{1/2}\,\\
 &\le&
 C\,
 \left\| \sup_{i\le N}\left\|t^{8,k}_{i,N}\right\|_\infty\right\|_2
 \,,\quad k=2,3.
 \ee
 Finally, combining relations \reff{unifborn}-\reff{upperboundvarphi2}, Cauchy-Schwarz inequality and \reff{HypDoubleNh},
 we get
 \b*
 \left\| \sup_{i\le N}\left\|t^{8,2}_{i,N}\right\|_\infty\right\|_2
 = \PtO\left(\frac{1}{Nh^{d+n+1}}\right)\,,
 \e*
 and
 \b*
 \left\| \sup_{i\le N}\left\|t^{8,3}_{i,N}\right\|_\infty\right\|_2
 = \GdO\left(\frac{(\ln N)^3}{Nh^{d+n+1}\sqrt{Nh^{d+n}}}\right)
 = \PtO\left(\frac{1}{Nh^{d+n+1}}\right)\,.
 \e*
 Therefore \reff{estimt81n} and \reff{estimestim} lead to the expected equivalent for $\Esp{T^8_N}$.

\no 2. We now study the variance of $T^8_N$. We first notice that
the Cauchy-Schwarz inequality and \reff{HypDoubleNh} lead to
 \b*
 Var\left[T^{8}_N\right]
 \;\le\;
 C\,
 \left\| \sup_{i\le
 N}\left\|t^{8}_{i,N}\right\|_\infty^4\right\|_4^{2}
 \e*
 But, using again Cauchy-Schwarz inequality and relations \reff{HypDoubleNh}, \reff{unifborn}, \reff{upperboundvarphi2} and
 \reff{estimatepollardmoment3}, we deduce that
 \b*
  \Var\left(T^8_N\right)
  &=&
  \GdO\left(\frac{\ln^4N}{N^2h^{2d+2n+2}}\right)
  =
  \PtO\left(\frac{1}{Nh^{d+2}}\right)\,.
  \e*
 \ep

\begin{Lemma}\label{lemT9}
$\E[T^9_N]$ $=$ $\GdO(N^{-1}h^{-d-n})$ and $\Var(T^9_N)\;=\;{\rm
o}(N^{-1}h^{-d-2})$\,.
\end{Lemma}

\proof It can be easily checked that $T^9_N$ can be dealt as $T^8_N$
and, following the lines of the proof of Lemma \ref{lemT8}, we
obtain the announced result.

\subsection{Asymptotic bias and variance}

This section is devoted to the proof of Proposition \ref{proptildebeta} characterizing the asymptotic bias and variance of the double kernel based estimator $\tilde \beta_N$.\\

\no{\bf Proof of Proposition \ref{proptildebeta}.}
We split the proof in two steps.\\
 1. We first derive the expectation of $\tilde\beta_N$.

Notice that $T^1_N$ $=$ $\bar\beta_N$ as defined in~\reff{barbetaN}
which satisfies\b* \E\left[\bar\beta_N\right] &=& \frac{1}{\ell(0)}
\int \phi(z)K(l)s(\lambda^0 - hl,z) \varphi(\lambda^0- hl,z)\,
      dt\,dz\,.
 \e*
The regularity of the function $s\varphi$ given by assumption {\rm
R} enables us to derive \be\label{ExpT1}\E[T^1_N]-\beta \sim
\frac{h^p}{\ell(0)} \int \,\xi^p_K\left[\,\ell f_\lambda
\right](\lambda^0,z) \phi(z)\,dz\,.\ee

Using remark \ref{remRestexpansionvarphi}, we deduce from
(\ref{expansionvarphil}) that we have \b*\label{ExpT2} \E[T^2_N] &=&
\frac{h^p}{\ell(0)} \int\,\xi^p_K\left[\varphil\right](\lambda^0,z)
\phi(z)\,dz\,+\,\frac{h^q}{\ell(0)}
\int\,\xi^q_H\left[\varphil\right](\lambda^0,z) \phi(z)\,dz\,+{\rm
o}(h^{p\wedge q})\,.\e*

We now rewrite $t^3_{i,N}$ as the sum of
 \b*
 t^{3,1}_{i,N}\,:=\, \frac{(\varphi-\bar\varphi)\,\varphil}{\varphi^2}
 \;\;\mbox{ and }\;\;
 t^{3,2}_{i,N}\,:=\,\frac{(\bar\varphi^\delta-\bar\varphi)\,\varphil}{\varphi^2} \,,\;\;
 \e*
 and study separately the corresponding $T^{3,1}_{N}$ and
 $T^{3,2}_{N}$. From \reff{expansionvarphi}, we derive
 \b* \E[T^{3,1}_N]\,=\,
-\,\frac{h^p}{\ell(0)} \int \,\frac{\varphil\xi^p_K\left[\varphi
\right]}{\varphi} (\lambda^0,z) \phi(z)\,dz-\,\frac{h^q}{\ell(0)}
\int \,\frac{\varphil\xi^q_H\left[\varphi \right]}{\varphi}
(\lambda^0,z) \phi(z)\,dz+\PtO(h^{p\wedge q})\,,
 \e*
 and we directly deduce from \reff{HypDoubleNh} and
 \reff{unifborn00} that $\E[T^{3,2}_N]=\PtO(h^{p\wedge q})$.

Note that
$$t^4_{i,N}=\frac{(\varphi-\bar\varphi^\delta)^2\varphil}{\varphi^2\bar\varphi^\delta}+\frac{(\bar\varphil-\varphil)(\varphi-\bar\varphi^\delta)}{\varphi\bar\varphi^\delta}.$$
Then, using \reff{unifborn}, \reff{unifborn2},
\reff{upperboundvarphi1} and \reff{upperboundvarphi2}, we derive
$||t^4_{i,N}||_{\infty} \,=\, \PtO\left(h^{p \wedge q}\right)$ and
Lemma \ref{lemExpectation} leads to $\,\E(T^4_N)=\PtO(h^{p \wedge
q})\,.$

From Lemmas \ref{lemT5}, \ref{lemT6} and \ref{lemT7}, we have
$\E(T^j_N)=0$ for $j=5\ldots 7$ and Lemma  \ref{lemT8} gives
 \b*
 \E\left[T^8_N\right]
 &\sim&
 \frac{h^{-d-n-1}}{\ell(0)N}
 \int\frac{\phi(z)}{\varphi(\lambda^0,z)}K(l_2-l_1)K(l_1)\nabla K(l_1)
 H^2(v) \, dl_1\, dv \, dl_2\, dz\,.
 \e*

Finally, Lemma \ref{lemT9} tells us $\E[T^9_N]$ $=$ ${\rm
o}(N^{-1}h^{-d-n-1})$.

We then obtain $\E[\tilde\beta_N]$ by summing up the $\E[T^j_N]$ for
$j=1,\ldots, 9$.\\

 2. We then analyze the variance of $\tilde\beta_N$.
For any $j=1,\ldots, 4$, expressions \reff{unifborn},
\reff{unifborn2}, \reff{upperboundvarphi1} and
\reff{upperboundvarphi2} imply $\,||t^j_N||_{\infty} \,=\, {\rm
O}\left(1\right)\,$. Then, Lemma \ref{lemVariance} leads to
$$\Var(T^j_N)=\PtO(N^{-1}h^{-d-2}) \quad \mbox{for every}\,
j=1,\ldots, 4\,.$$ From Lemma \ref{lemT5}, we get
\be\label{VarTot}\Var(T^5_N)\sim \frac{1}{\ell(0)Nh^{d+2}} \int
\phi^2(z)\left\{\int K(l_2-l_1) \nabla K(l_1) dl_1\right\}^\otimes
f(\lambda^0,z) \, dz \, dl_2\,.
 \ee
Indeed, Lemmas \ref{lemT6} to \ref{lemT9} imply also
$$\Var(T^j_N)=\PtO(N^{-1}h^{-d-2}) \quad \mbox{for every}\,
j=5,\ldots, 9\,.$$ Hence, $\,{\rm Cov}(T^j_N,T^k_N)={\rm
o}(N^{-1}h^{-d-2})$ unless $j=k=5$ and $\Var(\tilde\beta_N)$ is
given by expression \reff{VarTot}. \ep

\subsection{Central limit theorem}

This section is devoted to the proof of Theorem \ref{thmtildebeta},
which provides a central limit theorem for the double kernel based estimator $\tilde \beta_N$.\\

\no{\bf Proof of Proposition \ref{proptildebeta}.} As we saw in the
proof of Proposition \ref{proptildebeta}, the variance of
$\tilde\beta_N$ is given by the variance of \b*T^{5,2}_N &=&
\frac{h^{-2d-n-1}}{ \ell(0)\,N}\sum_{i} b(\Lambda_i,Z_i)
\,,\e*\vspace{-0.5cm}
 \b*
 \mbox{where}\qquad b(\lambda,z)
 &:=&
 h^{d+n} \int  \phi(z+hv)\;
              K\left(\frac{\lambda^0-\lambda}{h}-l\right)
              \nabla K (l)
              H (v) \,dl\,dv\\
 &-&
 h^{n+1} \int  \phi(z)\;\bar \varphil(\lambda^0 - hl,z) K (l) \,dl.
 \e*

As in the proofs of Theorems 4.1 or 4.2 in \cite{ours}, using
Kolmogorov's condition with the fourth moment of $b$ and the
Cramer-Wold device, we derive that $T^{5,2}_N$ is asymptotically
normal. We then finally deduce that
 \b*
 \sqrt{Nh^{d+2}} \left(\tilde\beta_N - \E[\tilde\beta_N] \right)  &\mathop{\stackrel{law}{\longrightarrow}}
 \limits_{N\rightarrow\infty}&
 \mathcal{N} \left(0,\tilde\Sigma \right)\;.
 \e*
 Under the additional condition $Nh^{d+2+2(p\wedge q)}\rightarrow 0$, we conclude the proof denoting that the bias vanishes in the previous expression.
 \ep

\end{document}